\documentclass[sigconf, nonacm]{acmart}
\usepackage{graphicx}
\usepackage{siunitx}

\AtBeginDocument{%
  }
\usepackage{balance} 

\begin{document}

\title{Preliminary Characterization of Bio-inspired Dog-Nose Sampler for Aerosol Detection}

\author{Yahya Naveed}
\affiliation{%
  \institution{University of Michigan}
  \city{Ann Arbor}
  \country{Michigan}}
\email{yahyanav@umich.edu}

\author{Julia Gersey}
\affiliation{%
  \institution{University of Michigan}
  \city{Ann Arbor}
  \country{Michigan}}
\email{gersey@umich.edu}

\author{Pei Zhang}
\affiliation{%
  \institution{University of Michigan}
  \city{Ann Arbor}
  \country{Michigan}}
\email{peizhang@umich.edu}
\renewcommand{\shortauthors}{Naveed et al.}

\begin{abstract}
Before aerosols can be sensed, sampling technologies must capture the particulate matter of interest. To that end, for systems deployed in open environments where the location of the aerosol is unknown, extending the reach of the sampler could lessen the precision required in sensor placement or reduce the number of sensors required for full spatial coverage. Inspired by the sensitivity of the canine olfactory system, this paper presents a rudimentary sampler that mimics the air flow of a dog's nose. The design consists of speed-controlled inhalation jets, as well as exhalation jets that are angled down and to the side. We tested this design on volatile organic compounds (VOC) in a small number of scenarios to validate the concept and understand how the system behaves. We show that in preliminary testing this dog-nose setup provides improvements over passive and solely inhalation sensing. 
\end{abstract}

\maketitle

\section{Introduction}
Aerosol detection impacts many fields such as air pollution monitoring, bomb detection, and viral tracing. Many of these fields are moving towards the utilization of smaller sensor nodes to track particulate matter in real time \cite{yu_real-time_2022, agarwal_modulo_2020, balid_development_2017, banach_new_2020, concas_low-cost_2021, egodagama_air_2021, gersey_pilot_2023, 10.1145/3560905.3568071, shi2025phy, chen2018pga, chen2016hap, munera_iot-based_2021, huffman_real-time_2020, puthussery_real-time_2023, gao2016mosaic, devarakonda2013real, xu2014gotcha, xu2016gotcha, liu2024mobiair, chen2020adaptive, liu2022fine}. However, accurate detection of aerosols in real-world environments is dependent on the ability to capture particulate matter effectively. Oftentimes, the origin of aerosols is unknown, making it challenging to ensure that they are being sensed in the environment. Extending the reach of sampling technologies to capture larger distances could mitigate this challenge. Furthermore, sampling technologies that can increase the volume of aerosols detected ease sensitivity requirements for the sensor and are therefore valuable for improving system robustness and efficiency in complex or dynamic environments.

Even as advances in sensor technology have continued, with many biosensors moving towards smaller, "system-on-chip" designs to enable real-time detection, limited work has been done on improving sampling technologies to follow this trend \cite{yahya_paper_2025, breshears_biosensor_2022, hong_gentle_2016, nam_air_2024, qiu_dual-functional_2020, seo_rapid_2020, stapf_membrane-based_2024, xiong_efficient_2021}. Nevertheless, preliminary work has explored using bio-inspired designs based on the dog nose to improve both sensing and sampling due to the exceptional smelling capabilities of a dog \cite{guest_feasibility_2021, staymates_biomimetic_2016}. For sampling technologies, the directed airflow of a dog's nose while exhaling is of particular interest because of its unique properties. Namely, instead of forcing air back through the same passage as inhalation, the air is expelled out at an angle (down and to the side). 

Inspired by the canine's olfactory system, this work presents the development of a rudimentary aerosol sampler that mimics aspects of a dog's sniffing behavior. The device incorporates speed-controlled inhalation jets and angled exhalation jets to actively manage airflow near the sampling region. We validated the design using volatile organic compounds (VOCs) across a limited set of controlled scenarios. Preliminary results revalidate that the dog-nose-inspired approach offers improved sampling performance compared to passive and inhalation-based systems at distance. Using the modular nature of the inhalation jets, we also characterize the response to differences in air intake speed at a fixed distance from the VOC source. We hope that these results support its potential for future aerosol detection applications.

\section{System Design}
The design was inspired by the airflow of a dog nose. While the human respiratory system directs inhaled and exhaled air in the same direction, a dog can direct exhaled air in a different direction, namely down and to the side. For a dog, this redirection is possible due to the slits seen on the nostril. Following this idea, we developed a system that decouples the inhalation and exhalation vents of our sampler in order to mimic this phenomenon. 
The system design can be broken down into two main categories: the chamber, including the dog "nostrils," and the control circuitry for the airflow motors. 

\subsection{Airflow Chamber}
\begin{figure}[!htb]
    \centering
    \includegraphics[width=0.8\linewidth]{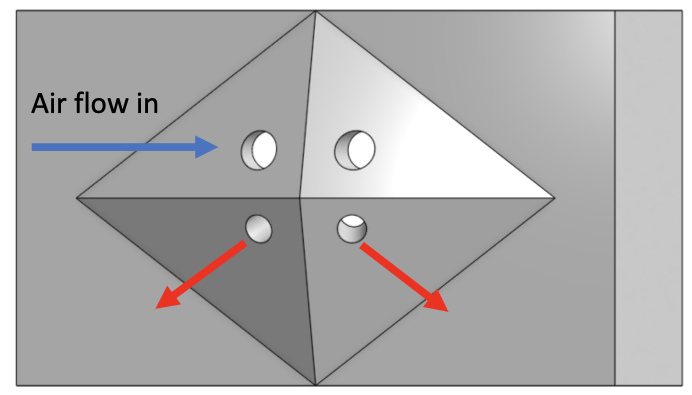}
    \caption{3D CAD Model of Dog Nose Design}
    \label{fig:3dmodel}
\end{figure}

The airflow chamber is where the sampled air is held. As seen in Figure \ref{fig:3dmodel}, the design's air intake vents are modeled to behave similarly to a dog's nose, and it uses a pyramid-shaped "snout" to achieve this goal. The top two holes are designated for "inhalation" and are drilled orthogonal to the base of the pyramid while the bottom two holes are for "exhalation" and are drilled along the bottom faces of the pyramid. The distance between the "nostrils" was 19.32mm. Changing the length of the snout (the height of the pyramid, currently 1.27cm) can also allow for different angles of exhalation, which has not yet been characterized. 

Tubing was used to isolate the expelled air from the rest of the system, in the case the system was used to detect viral matter. Therefore, the radius of the inhalation and exhalation holes differ, as the tubing matched the motors that would actuate the airflow in either direction. The inhalation hole size was 8mm diameter distance whereas the exhalation tubing size was 5.5mm. 

\begin{figure}[!htb]
    \centering
    \includegraphics[width=0.8\linewidth]{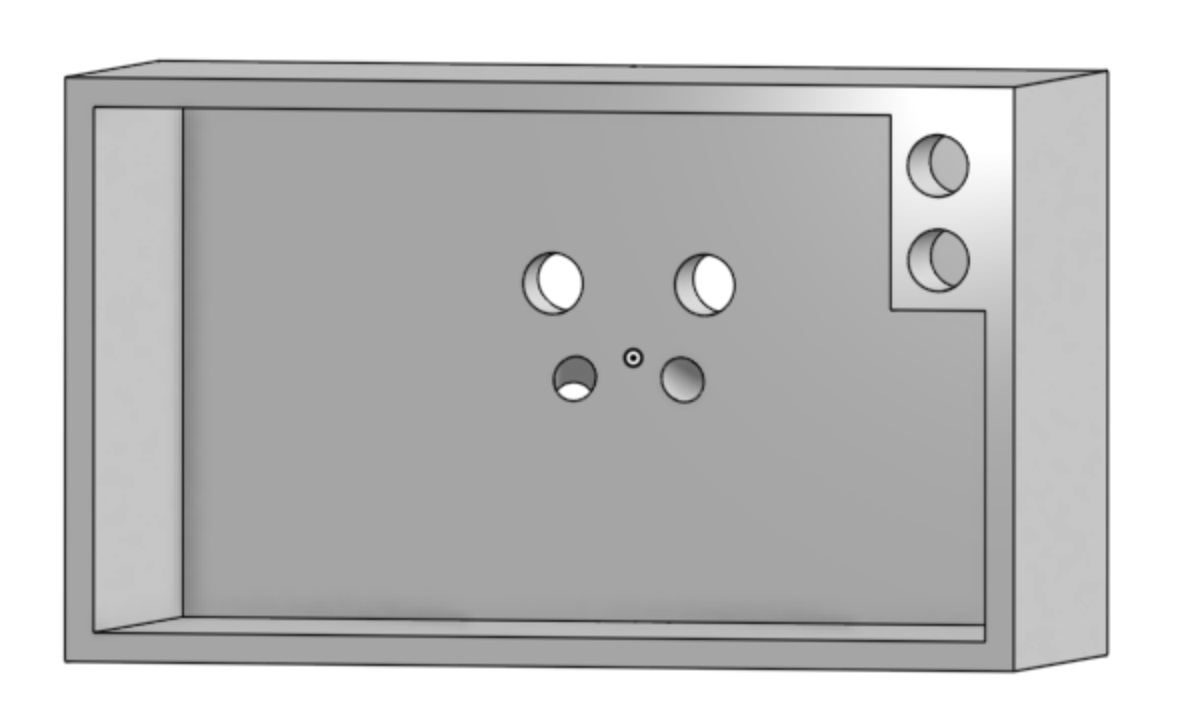}
    \caption{Back of 3D CAD Model of Dog Nose Design}
    \label{fig:3dmodel_back}
\end{figure}

Figure \ref{fig:3dmodel_back} shows the back of the chamber designed for aerosol capture. After several design iterations, a dedicated mount for the inhalation motor tubing was incorporated to ensure efficient airflow into the chamber. To date, the sensor has been attached to the back opening of the chamber and sealed with adhesive. Future iterations would design for compatibility with specific sensors to limit VOC exposure from outside sources, such as adhesives. The final model was printed in PLA, selected because it exhibited low levels of volatile organic compound (VOC) emissions following fabrication.

\subsection{Motor Control Circuit}
To enable airflow characterization beyond the exhalation angle, the 12V inhalation motor was integrated into a PWM-actuated control system. A Seeed Studio ESP32 microcontroller was used to set the PWM output values. The exhalation motor was similarly connected to the PWM system; however, it required a voltage step-down to 3.5V using a buck converter. Due to this constraint, PWM was used solely for power delivery to the exhalation motor, without modulation.

Implementing PWM control provided the opportunity for further exploration into understanding the system’s response to various airflow patterns. In addition to limiting power to the motors and reducing airflow, PWM enables the motors to be pulsated on and off to simulate breathing cycles. Due to limitations in DC motor shutoff response times, the fastest pulsing period achieved was 20 seconds. This period is much slower than a dog's breathing, which is approximately 0.2 seconds (5 Hz) \cite{staymates_biomimetic_2016}.

\section{Test Setup}
\begin{figure}[!htb]
    \centering
    \includegraphics[width=0.8\linewidth]{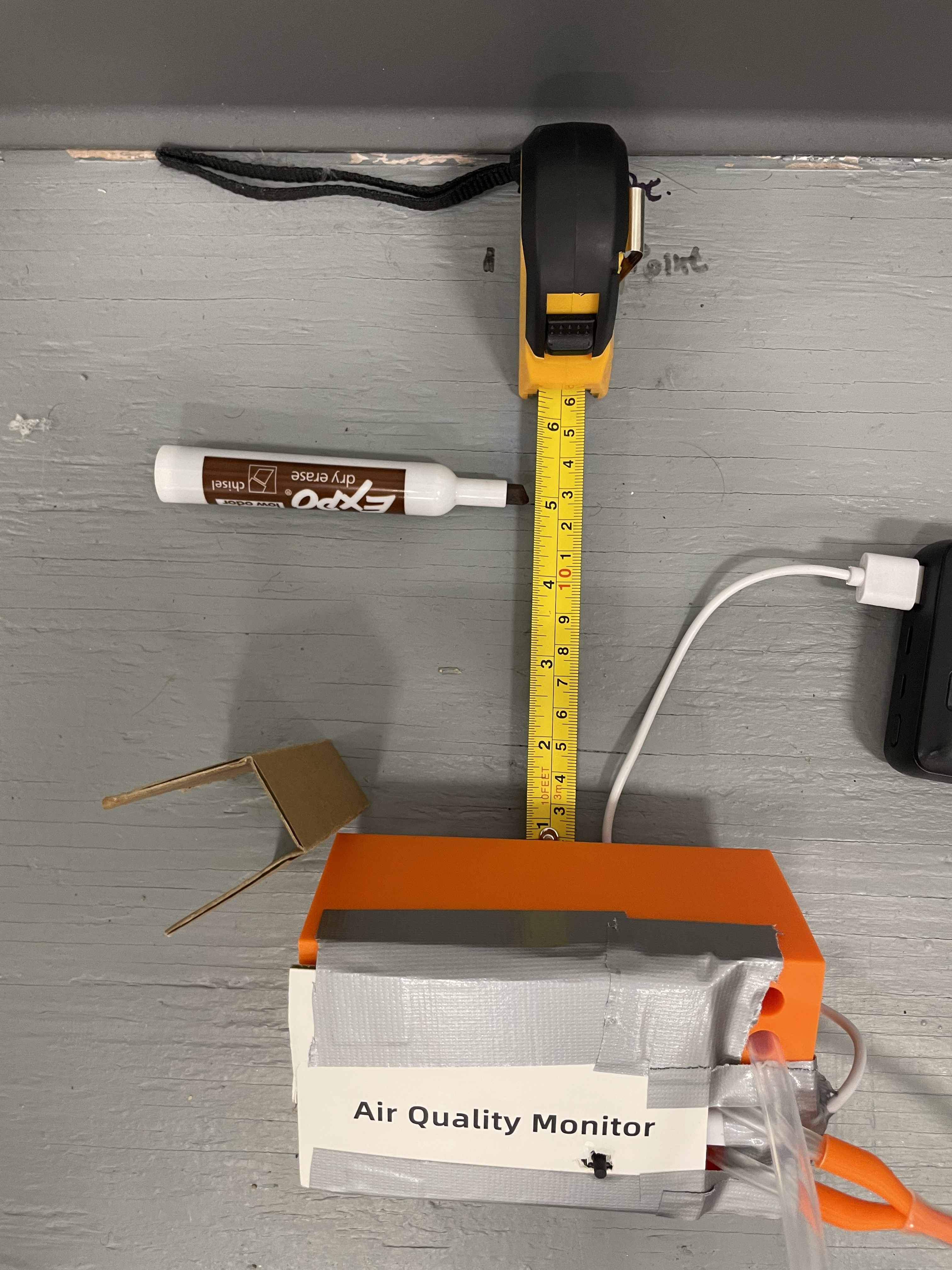}
    \caption{Test Setup Including Orientation and Source Location}
    \label{fig:testsetup}
\end{figure}

Figure \ref{fig:testsetup} shows the general test setup for the system. The air chamber is facing downward, with the sensor attached on the back. The VOC source, an uncapped dry erase marker, is set five inches above the top of the module, and it is uncapped only after data capture begins. The height from the ground of the module is varied in multiples of 1.27cm, the length of the module "snout." To raise the module, cardboard stilts of the required length were inserted underneath, along the shorter edges of the system, as to be minimally invasive to the airflow. 

\begin{figure}[!htb]
    \centering
    \includegraphics[width=0.8\linewidth]{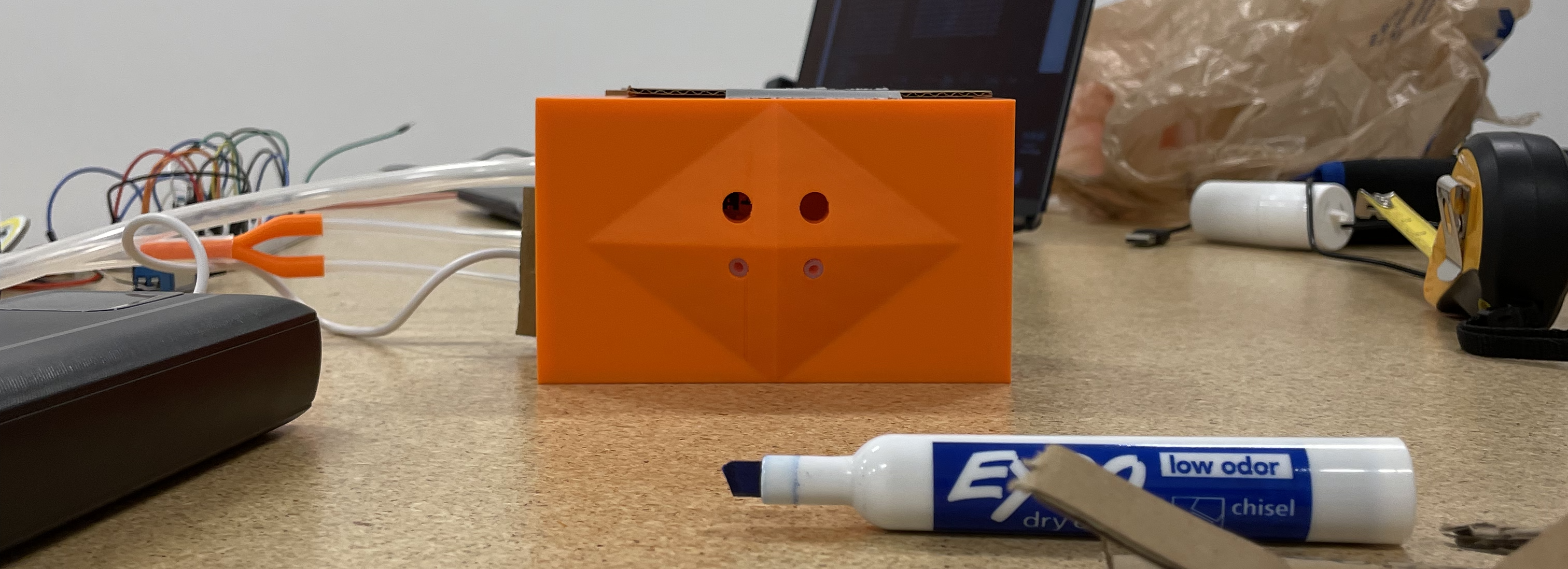}
    \caption{Test Setup For 90 Degree Orientation}
    \label{fig:testsetup90}
\end{figure}

A small number of tests were also performed with the module rotated 90 degrees. For these tests, the module was placed on the surface with the snout facing the VOC source, again 5 inches away, as seen in Figure \ref{fig:testsetup90}. 

\section{Results}
Using this test setup, the results indicate that the dog-nose setup does provide benefit compared traditional sampling - for both passive sampling and solely inhalation into the sampler. The first tests revalidate previous work for a dog-nose-based system but additional tests are done to further understand the best usage of the sampler \cite{staymates_biomimetic_2016}. The sections also discuss future tests that may be done to continue this work based on the given results. 

\subsection{Baseline Analysis}
\begin{figure}
    \centering
    \includegraphics[width=1\linewidth]{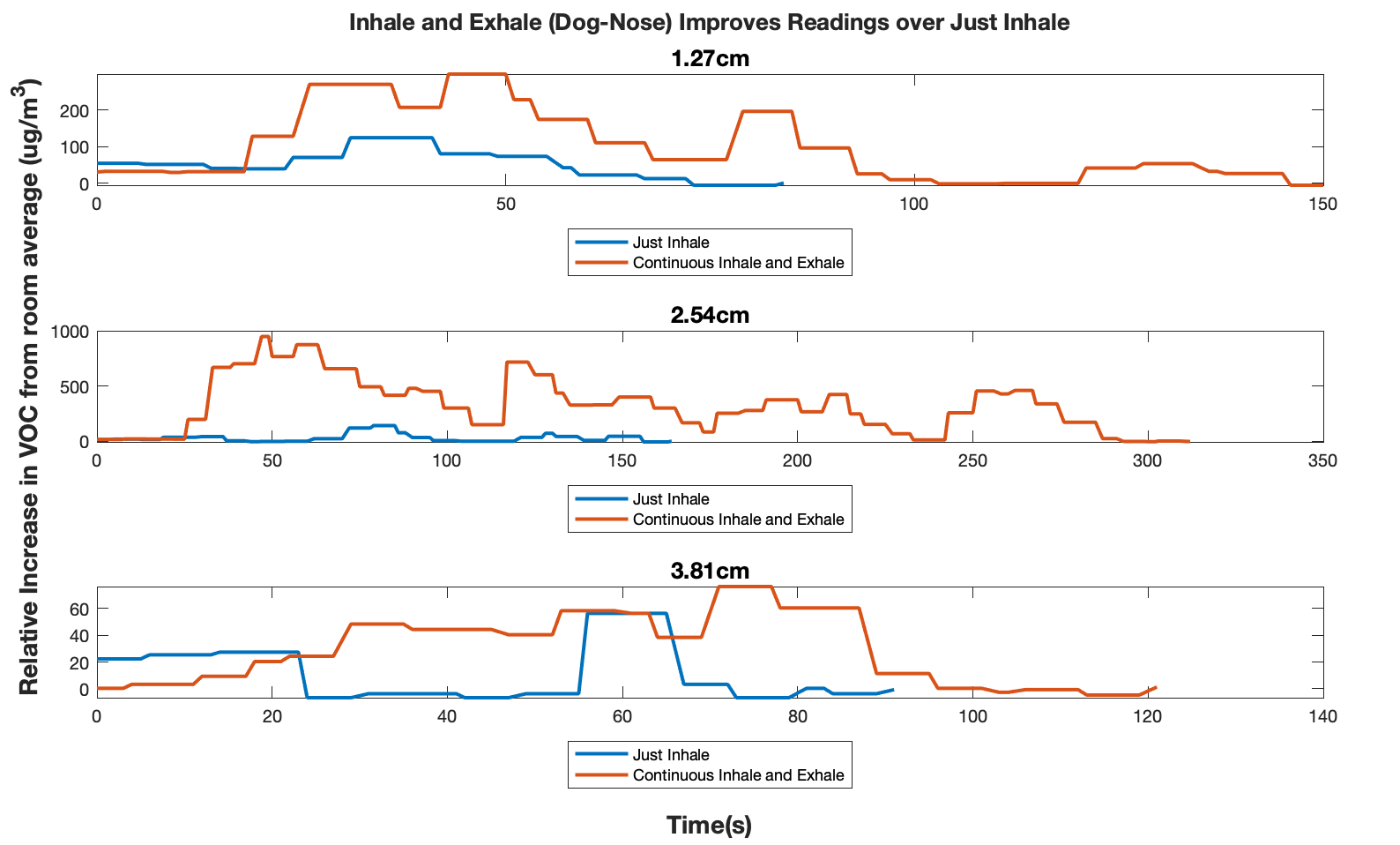}
    \caption{Baseline Analysis of Inhalation vs Continuous Inhale and Exhale (Dog-Nose) at Varying System Heights}
    \label{fig:baseline}
\end{figure}

\begin{figure}
    \centering
    \includegraphics[width=1\linewidth]{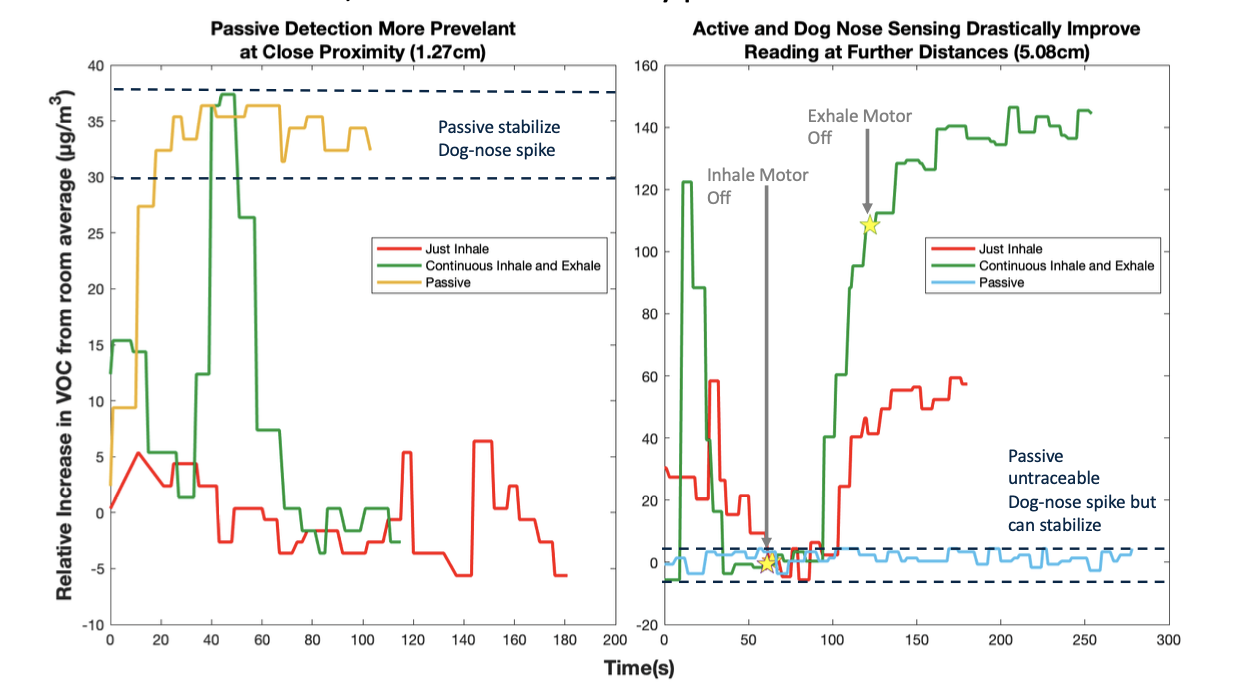}
    \caption{Baseline Analysis Including Passive, Inhalation, and Continuous Inhalation and Exhalation (Dog-Nose) at 1.27cm and 5.08cm}
    \label{fig:passivevactive}
\end{figure}

The system was first characterized for performance compared to a baseline (continuous inhalation without exhalation and fully passive reception of VOCs). Figure \ref{fig:baseline} shows the results of the dog-nose system compared to continuous inhalation, where it is seen that the dog-nose is able to capture a larger amount of VOCs as well as detect these VOCs for longer. Figure \ref{fig:passivevactive} shows an interesting result in which passive detection outperforms either airflow scheme when the system is positioned very close to the surface. In this case, where the tip of the dog-nose is touching the surface, passive detection reaches a similar peak as the active system but maintains detection for a longer period. However, once the system is elevated from its surface, passive detection does not yield notable results from room-average VOC levels, whereas both inhalation and dog-nose airflow schemes capture better results. Again, the dog-nose detects a higher peak in VOC concentration than just inhalation. 

The graph on the right of Figure \ref{fig:passivevactive} shows the spike in the VOC count after the inhalation motors of each system were turned off, and this trend would continue to appear in other tests. Further investigation must be done to isolate for the exact cause of this phenomenon, but later tests will show that the inhalation mechanism causes this result irrespective of the presence of a VOC source. 

\subsection{Varied Inhalation Speed}
\begin{figure}
    \centering
    \includegraphics[width=1.2\linewidth]{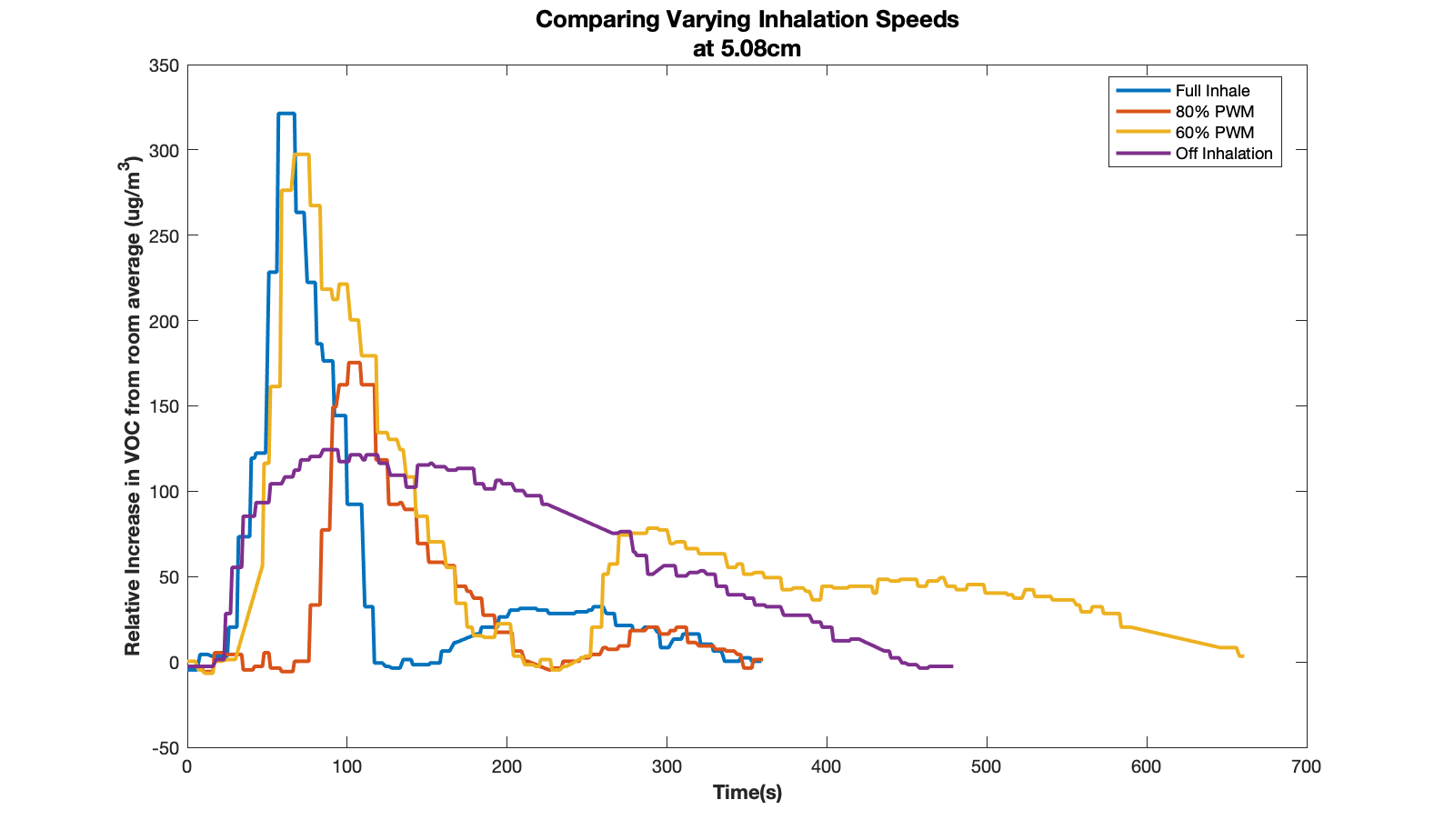}
    \caption{Response with Varying Inhalation Air Speeds via PWM and Constant Exhalation}
    \label{fig:speeds}
\end{figure}

The strength of the airflow into the chamber was varied, and the corresponding VOC concentrations were measured, as shown in Figure \ref{fig:speeds}. While full-power inhalation produced the highest peak VOC concentration at approximately 325µg/m$^3$, reducing the motor speed to 60\% PWM still resulted in a comparable peak of around 300µg/m$^3$. Furthermore, 60\% PWM maintained elevated VOC levels for a longer duration. Interestingly, operating the motor at 80\% duty cycle was less effective than both 60\% and 100\%, suggesting a possible bimodal relationship in achieving peak VOC concentrations. Unfortunately, due to motor limitations, higher airflow rates could not be tested to validate this idea. In general, the duration of detectable VOC levels appeared to be inversely proportional to motor strength. This suggests that, after VOCs enter the chamber, continuous high-speed airflow may draw them away from the sensor and toward the tubing, even possibly completely out of the chamber.

\subsection{90 Degree Orientation}
\begin{figure}[!htb]
    \centering
    \includegraphics[width=1\linewidth]{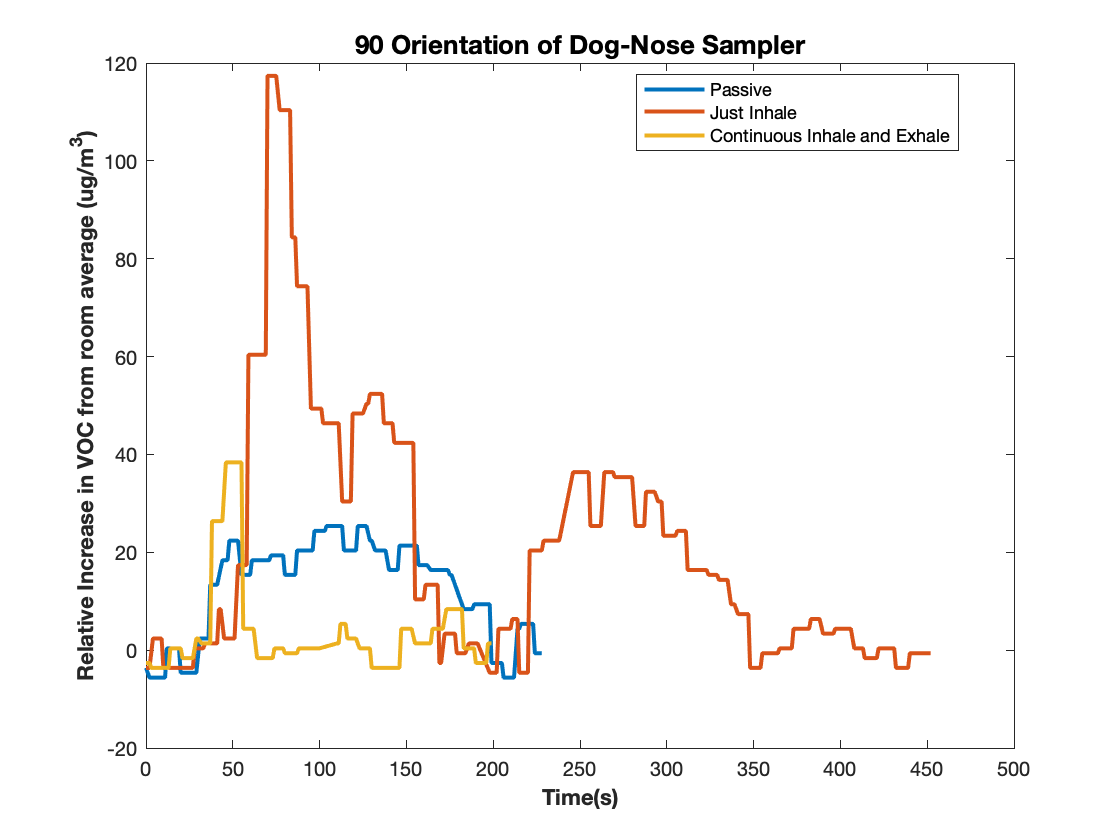}
    \caption{Captured Response for Tests using the 90-degree Orientation of the Dog Nose System}
    \label{fig:ninety}
\end{figure}

A 90 degree orientation was also tested using this setup, shown in Figure \ref{fig:testsetup90}. In these tests, shown in Figure \ref{fig:ninety}, running just the inhalation motor provided the best response; although, this response contained a lower maximum VOC count than the dog-nose tests at 5.08cm in the face-down setup in Figure \ref{fig:passivevactive} and Figure \ref{fig:speeds}. It is probable given this result and the result from the close proximity test in Figure \ref{fig:passivevactive} that the exhalation component of the dog-nose sampler requires the displacement of a large air space to work effectively, which close proximity to a solid surface prohibits. 

A more comprehensive set of angle tests should be conducted to determine the optimal positioning of the dog-nose relative to the source. Additionally, it may be valuable to revisit the 90-degree orientation tests by placing the dog-nose at this angle near the edge of a surface, allowing for greater displacement of air into open space.

\subsection{Other Breathing Schemes}
\begin{figure}
    \centering
    \includegraphics[width=0.8\linewidth]{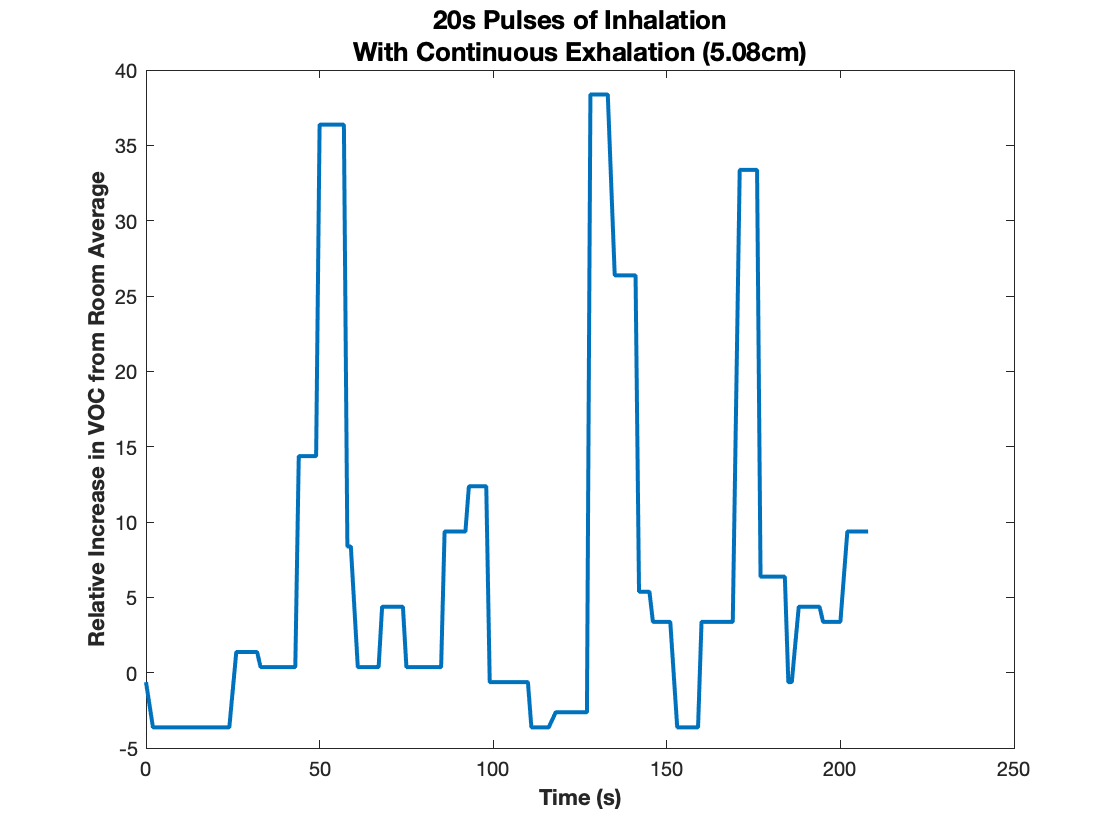}
    \caption{Response with 20s Pulses}
    \label{fig:pulse}
\end{figure}

Up to this point, most tests have ran the airflow motors continuously or not at all; however, with the control scheme implemented, there lies potential for exploration into other breathing patterns, such as mimicking a dog's breathing pattern - quick successions of inhalation followed by exhalation. With the motor turn-on limitations previously discussed, the quickest time period achieved in operating the motors was 20 seconds. At this speed, the inhalation motor was pulsated, while the exhalation motor was continuously ran, and the results are shown in Figure \ref{fig:pulse}. At a distance of 5.08cm off the ground, there is little difference between the concentration obtained from this measurement and the solely inhalation from Figure \ref{fig:passivevactive}, but this method did provide cyclical readings of that concentration. Repeating this test at a quicker frequencies and alternation between inhalation and exhalation motor turn on should be considered. 

\section{System Improvements}
\subsection{Post-Inhalation VOC Peaking}
As seen in previous figures, after the inhalation motor is turned off, the VOC count begins to peak once more. Although the exact cause of this phenomenon is uncertain, there are two leading possibilities currently. The first is that the inhalation causes VOCs to enter the chamber but are quickly removed before detection. This is possible due to where the tubing is connected and the continuous running of the motor; with the tubing being off to the side, VOCs will be channeled in that direction and through the tube. Once the motor is stalled, the VOCs are able to settle within the chamber and the sensor is able to read them. The second possibility is that the tubing material, being a plastic itself, introduces some VOCs into the system, and these are only detectable once the motor is switched off. It is most likely that the inhalation subsystem causes this phenomenon, since it actively brings VOCs into the chamber to be detected; the exhalation air system is isolated from the chamber using its own tubing. Additionally, from Figure \ref{fig:speeds}, when inhalation is turned off but the exhalation motors remain on, no such peaking was detected. 

\begin{figure}
    \centering
    \includegraphics[width=0.8\linewidth]{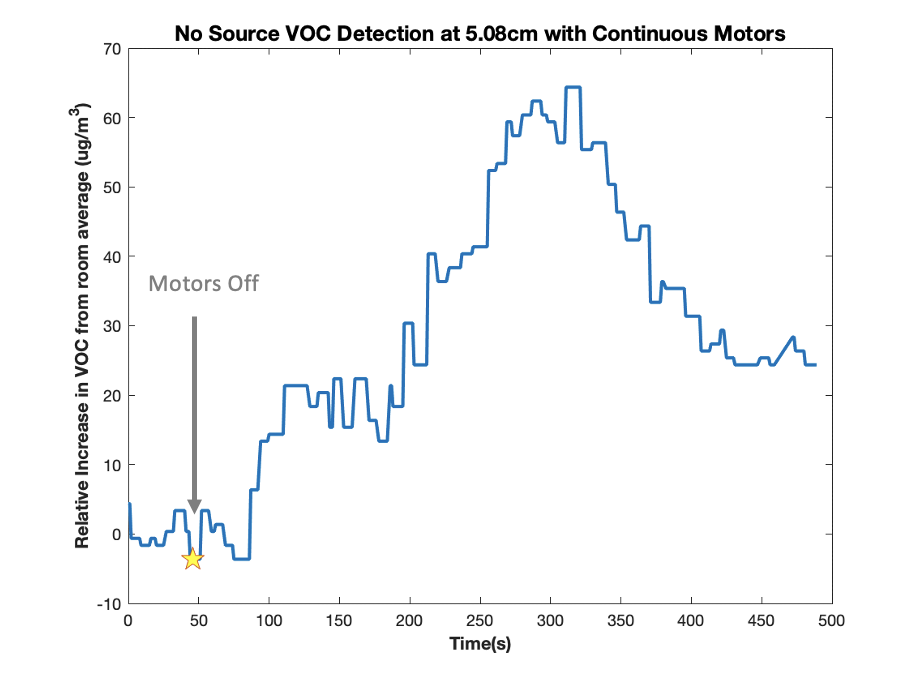}
    \caption{No Source Still Induces VOC Peaking After Motor Turn Off}
    \label{fig:nosource}
\end{figure}

Even without a source present, the post-inhalation VOC peaking is present, as shown in Figure \ref{fig:nosource}. 

\subsection{Choice of Materials and Sensor}
This first prototype shows much promise when it comes to improving sampling volume of aerosols. However, this version incurred some problems while testing. Some time was taken between validation testing of the idea and conducting further unique tests to characterize the design. During this time, it was found that 3D-printed materials remain porous, and VOCs can become embedded into these pores. Compared to the ambient, the baseline VOC count within the chamber of the first print continually read much higher, which would skew results. Therefore, a new model was printed and used to finish preliminary characterization. In future studies, it may help to use non-plastic materials or to utilize a sensor that can capture other aerosols, such as PM2.5, to see the benefit of this scheme for different aerosol sizes.


\balance
\bibliographystyle{ACM-Reference-Format}
\bibliography{main}

\appendix

\end{document}